\documentstyle[aps,prb,preprint]{revtex}
\begin{document}

\draft

\title{Exchange
Instabilities in Semiconductor Double Quantum Well Systems }

\author{Lian Zheng, M. W. Ortalano, and S. Das Sarma}
\address{Department of Physics, University of Maryland, College Park, Maryland
20742-4111}

\date{\today}

\maketitle

\begin{abstract}
We consider various exchange-driven electronic
instabilities in semiconductor double-layer systems 
in the absence of any external magnetic field.
We establish that there is no exchange-driven bilayer to monolayer charge 
transfer instability in the double-layer systems.
We show that, within the unrestricted Hartree-Fock approximation,
the low density stable phase (even in the absence of any
interlayer tunneling) is a quantum ``pseudospin rotated'' spontaneous
interlayer phase coherent spin-polarized 
symmetric state rather than 
the classical Ising-like
charge-transfer
phase. 
The U(1) symmetry of the double quantum well system is broken spontaneously
at this low density quantum phase transition, and the layer density develops 
quantum fluctuations even in the absence of any interlayer tunneling.
The phase diagram for the double quantum well system is calculated 
in the carrier density--layer separation space, and the possibility of
experimentally observing various quantum phases 
is discussed.
The situation in
the presence of an external electric field is investigated in some detail using
the spin-polarized-local-density-approximation-based self-consistent technique
and good agreement with existing experimental results is obtained.

\end{abstract}




\section{Introduction}
\indent Exchange driven instabilities in an electron gas have been a subject of
long standing
interest \cite{HER} in condensed matter physics dating back to 1929 when Bloch
first
pointed out \cite{BLOCH} that a low
density free electron gas may undergo a spontaneous spin polarization
transition to a ferromagnetic
state by virtue of the dominance of exchange energy over kinetic energy at
low enough electron density. A number of possible exchange instabilities has
been proposed and extensively studied theoretically \cite{HER} in three
dimensional
 electron systems including ferromagnetism, antiferromagnetism, and various
spin/charge texture phases. It is, however, unclear whether any such
exchange instability has ever been experimentally observed in a three
dimensional
free electron-like system. One problem is that the available three dimensional
free electron systems, namely alkali metals, have reasonably high effective
electron densities, making the normal paramagnetic ground state
energetically stable and exchange instabilities unlikely \cite{HER}. Recent
interest
in this subject has focused on the possibility of exchange instabilities
in two dimensional electron systems as occurring in artificially structured
semiconductor quantum wells, heterostructures, and superlattices. These
two dimensional electron systems, particularly the ones existing in
modulation doped GaAs-$\mbox{Al}_{x} \mbox{Ga}_{1-x}$As nanostructures,
offer several advantages over three dimensional electron systems ({\em e.g.},
metals, doped bulk semiconductors) in terms of a systematic study of
exchange-correlation effects. First, lower dimensionality typically
enhances interaction effects, making exchange instabilities more likely in
two dimensional electron systems. Second, the electron density can be
varied over (almost) two orders of magnitude in modulation doped two
dimensional systems (either by varying the modulation doping level and/or by
using suitable gates), thereby enabling one to tune the relative magnitude
of exchange-correlation effects. Third, these artificially structured two
dimensional systems can be made ultrapure (substantially reducing
disorder effects) because the ionized dopants are spatially separated from
the electron layer. Fourth, artificial structuring enables the
introduction of additional degrees of freedom into the problem, {\em e.g.}
separation between the layers in a bilayer system, which are
not available in purely two/three dimensional electron systems, thus
allowing the possibility of further tuning interaction effects.
Finally, and perhaps most importantly, the application of a strong external
magnetic field perpendicular to the two dimensional layer quenches the
kinetic energy of the system as the two dimensional electron gas gets
quantized into Landau levels, thereby increasing the importance of
electron-electron interaction effects. Because of these reasons as well as the
obvious
reason of substantial experimental and technological relevance, there has
been a great deal of recent interest in the possibility of interaction
({\em i.e.} exchange-correlation) induced exchange instabilities in two
dimensional systems. While much of this recent activity
\cite{SONG1,FERTIG1,COTE1,CHEN1,MACD1,MOON1,ZHENG,MACD2}
 focuses on the
situation in the presence of an external magnetic field, there has also been
considerable interest \cite{MACD2,RW,DATTA,SWI1,SDS1,KAT1,YING,PATEL} in the
possibility of
exchange
instabilities in two
dimensional electron gases in the absence of any external magnetic fields.
In this paper we theoretically investigate a specific zero magnetic field
exchange instability, namely a charge transfer instability, which has been
predicted to occur in semiconductor double quantum well systems
\cite{MACD2,RW,KAT1,YING,PATEL} under
suitable conditions.

The basic issue we study is quite simple. Consider a semiconductor double
quantum well structure ({\em e.g.} $\mbox{Al}_{x}
\mbox{Ga}_{1-x}$As-GaAs-$\mbox{Al}_{y} \mbox{Ga}_{1-y}$As-GaAs-$\mbox{Al}_{x}
\mbox{Ga}_{1-x}$As system) at
zero temperature which
has been modulation doped to produce a bilayer two dimensional electron system
(in the $x$-$y$ plane) with a layer separation $d$ (in the $z$ direction)
and a \underline{total} two dimensional electron density $2n$
(per unit area). Simple electrostatic considerations imply that the equilibrium
situation, which minimizes the Coulomb energy, is a classically symmetric
situation
with each quantum well equally populated with an electron density $n$.
(Quantum) Kinetic energy is also minimized by having equal populations of both
layers as this leads to a lower Fermi energy. Thus the n\"{a}ive expectation
(which, as we shall prove in this paper, turns out to be correct in this case)
is that the double quantum well system prefers a bilayer electron gas with
each layer equally populated with electrons. It has, however, been pointed
out \cite{MACD2,RW} that this simple picture may break down at low 
density and small
interlayer separation where there could be
a zero-temperature (quantum) phase transition from a bilayer to a monolayer
system driven entirely by exchange-correlation effects. This low density
bilayer to monolayer phase transition \cite{RW} is, in fact, an exchange
instability where at some low values of $n$, there is predicted to be a
spontaneous charge transfer from one layer to another, resulting in a
symmetry-broken monolayer phase where, instead of a bilayer electron system
with each layer having an electron density of $n$, all the electrons
reside in one layer with an electron density of $2n$.  This transition is
similar to the exchange-driven ferromagnetic spin polarization transition.
This conclusion on
the existence of a charge transfer instability in a double quantum well
system was reached in ref. \onlinecite{RW} by considering the competition among
the kinetic energy, the Coulomb (Hartree) charging
energy, and the exchange energy. Our goal is to investigate the problem
in the unrestricted Hartree-Fock approximation (HFA) by treating 
the layer index as
a fully quantum mechanical variable.
We conclude that there can be
no exchange-driven charge transfer instability  in a semiconductor
quantum well system under any conditions. 
The suggested charge transfer instability
is a feature of the restricted HFA 
where the layer index is treated as a classical Ising-like variable.
In the more general unrestricted HFA, there is an exchange driven instability 
towards a transition to a low density symmetric phase rather than the 
monolayer phase.

     A related issue we investigate connects with the recent experimental
search \cite{KAT1,YING,PATEL} to observe the predicted charge transfer
instability with
some of the papers \cite{KAT1} reporting experimental support for an abrupt
double-to-single-layer transition in a double quantum-well structure. These
experimental studies involve measurements of layer electron densities
[via low field Shubnikov-de Haas (SdH) oscillations] in a
double quantum well system
\underline{under the application of an external electric field}. The
applied electric field explicitly breaks the layer symmetry in the problem,
and the observed nonlinearity \cite{KAT1,YING,PATEL} in the layer
depopulation is a
direct
manifestation of the so-called exchange-correlation induced ``negative
compressibility''
effect \cite{EISEN}.
We study 
the layer/subband electron densities in the GaAs double quantum
well structures
in the presence of an applied electric field
within the self-consistent spin-polarized local-density-approximation, 
obtaining excellent agreement with the existing experimental
measurements \cite{KAT1,YING,PATEL}. The same self-consistent
approximation is used
to calculate
the phase diagram of the double quantum well system in the absence of any
external electric field and no stable monolayer electronic phase is found.

     The rest of this paper is organized as follows. In section II we 
investigate the phase diagram of a double quantum well 
structure in the electron
density ($n$)--layer separation ($d$) space
within the restricted HFA including effects of electron spin.
In section III we allow for the
possibility that the layer electron density is not required to be a good
quantum number \underline{even in the absence of interlayer tunneling}
and explicitly
include quantum
fluctuations in the layer density by considering 
symmetric quantum states which are
linear superpositions of electron states confined to different layers.
Such a ``pseudospin-rotated'' quantum state,
which involves no charge transfer, is
shown to always have a lower energy than the monolayer charge transfer
phase, 
establishing unambiguously that the
monolayer phase is \underline{not} energetically stable
in the HFA. In section IV we consider the
recent ``charge-transfer''
experiments in
double quantum well systems in the presence of external electric fields,
obtaining quantitative agreement between measured electron densities and
self-consistent spin-polarized local density calculations. We conclude with
a discussion
in section V.

\section{Restricted Hartree-Fock Approximation}
\label{RHFT}

 Consider a double quantum well system
where each electron
is in spin up/down and in layer left (or, layer 1)/right (or, layer 2)
 states (Fig. \ref{I}). In this
system there
are four possible (completely polarized or unpolarized) phases
which are denoted $\mbox{S}_{0}$ (equal population of
both
layer and spin components: the normal bilayer paramagnetic phase),
$\mbox{S}_{1}$ (equal population of both layers, but the electrons are spin
polarized
in each layer: the bilayer ferromagnetic phase), $\mbox{A}_{0}$ (equal
population of each spin component, but all the electrons are in a single layer:
the
monolayer paramagnetic phase), $\mbox{A}_{1}$ (the electrons are spin polarized
and reside only in one layer: the monolayer ferromagnetic phase).
Earlier work \cite{RW} did not explicitly consider the possibility of an
exchange-driven spin polarization transition (considering only paramagnetic
phases with equal populations of both up and down spins) and therefore included
only the possibility of $\mbox{S}_{0}$ and $\mbox{A}_{0}$ phases in their
restricted HFA of double quantum well charge transfer
instability. 
The fundamental principle underlying the
exchange instability is that exchange interaction prefers a spatially
antisymmetric wavefunction which, by keeping the electrons away from each
other,
optimizes the interaction energy. This can be accomplished equally effectively
by having a symmetric spin state ({\em i.e.} a spin polarized ferromagnetic
state)
and/or by having a symmetric layer state ({\em i.e.} a monolayer state), which
will
necessarily imply that the spatial part of the wavefunction is antisymmetric.
Thus, exchange should lead to a spin polarization ferromagnetic transition in
each layer as much as the bilayer to monolayer transition. In fact, the
exchange
driven
intralayer ferromagnetic transition ($\mbox{S}_{0} \rightarrow \mbox{S}_{1}$)
is more likely than the bilayer to monolayer transition ($\mbox{S}_{0}
\rightarrow \mbox{A}_{0}$)
because there is no Coulomb Hartree
energy to overcome in the spin polarization transition.
With this introduction to the possible spin/layer phases of the double
quantum well system, we discuss the HFA to the
ground state energy including only kinetic, exchange, and Hartree energy
contributions.
Following refs. \onlinecite{MACD2,RW} we model each electron layer as a
two-dimensional sheet of
zero thickness which then allows for a simple analytic calculation of the
ground state HFA energy per unit area, $E^{\rm HF}(n,d)$  as a
function of
the layer
separation $d$ and the electron density $n$.

\indent For a two layer system with $n_{i}$ and $m_{i}$ being respectively
the electron
density
and the spin polarization index/magnetization of layer $i$ ($i=1$ or 2),
the total energy per unit area within the
HFA is [here $m=(n_{\uparrow}-n_{\downarrow})/
(n_{\uparrow}+n_{\downarrow})$ is the spin polarization in a particular layer,
with $n_{\uparrow (\downarrow)}$ being the number density of spin up
(down) electrons in the layer]

\begin{eqnarray}
\label{eqn1}
E^{\rm HF}  & = &  \frac{e^{2}n_{1}}{2a^{*}} \left(
\frac{1+m_{1}^{2}}{r_{s1}^{2}}
- \frac{4\sqrt{2}}{3 \pi r_{s1}}\left((1+m_{1})^{3/2}+(1-m_{1})^{3/2} \right)
\right) \nonumber  \\
&  & + \frac{e^{2}n_{2}}{2a^{*}} \left( \frac{1+m_{2}^{2}}{r_{s2}^{2}} -
\frac{4\sqrt{2}}{3 \pi r_{s2}}\left((1+m_{2})^{3/2}+(1-m_{2})^{3/2} \right)
\right) \nonumber  \\
&  & + \frac{2 \pi e^{2} d}{\kappa_{barrier}} \left( \frac{n_{2}-n_{1}}{2}
\right)^{2}
\end {eqnarray}
where $r_{si}=1/(a^{\ast} \sqrt{ \pi n_{i}})$, with $a^{\ast}=\kappa
\hbar^{2}/m^{*} e^{2}$, where $\kappa=(\kappa_{well}+\kappa_{barrier})/2$ is
the
lattice dielectric constant, as the effective Bohr radius for the double
quantum well system. Note that in general $0 \le n_{i} \le 2n $ with the
constraint $n_{1}+n_{2}=2n$, and $0 \le |m_{i}| \le 1$. The various
contributions in
Eq. (\ref{eqn1}) for the HFA energy are
the kinetic energy (the two terms involving $r_{s}^{-2}$), the
exchange energy (the two terms involving $r_{s}^{-1}$), and the
electrostatic Hartree Coulomb energy associated with
charge transfer (the last term). 

\indent For the state $\mbox{S}_{0}$, $n_{1}=n_{2}=n$ and $m_{1}=m_{2}=0$.

\begin{equation}
\label{eqn2}
E^{\rm HF}   =   2  \left( \frac{1}{r_{s}^{2}} - \frac{8\sqrt{2}}{3 \pi r_{s}}
\right) \frac{e^{2}n}{2a^{*}}
\end{equation}

\indent For the state $\mbox{S}_{1}$, $n_{1}=n_{2}=n$ and $m_{1}=-m_{2}=1$.

\begin{equation}
\label{eqn3}
E^{\rm HF}  =   2  \left( \frac{2}{r_{s}^{2}} - \frac{16}{3 \pi r_{s}} \right)
\frac{e^{2}n}{2a^{*}}
\end{equation}

\indent For the state $\mbox{A}_{0}$, $n_{1}=0$, $n_{2}=2n$, and
$m_{1}=m_{2}=0$.

\begin{equation}
\label{eqn4}
E^{\rm HF} = \left( 2 \left( \frac{2}{r_{s}^{2}} - \frac{16}{3 \pi r_{s}}
\right) +
\frac{4d}{a} \frac{1}{r_{s}^{2}} \right) \frac{e^{2}n}{2a^{*}}
\end{equation}

\indent For the state $\mbox{A}_{1}$, $n_{1}=0$, $n_{2}=2n$, and $m_{1}=0$, and
$m_{2}=1$.

\begin{equation}
\label{eqn5}
E^{\rm HF} = \left( 2 \left( \frac{4}{r_{s}^{2}} - \frac{16 \sqrt{2}}{3 \pi
r_{s}}
\right) + \frac{4d}{a} \frac{1}{r_{s}^{2}} \right) \frac{e^{2}n}{2a^{*}}
\end{equation}

\noindent In Eqs. (\ref{eqn4}) and (\ref{eqn5}), $a=\kappa_{barrier} a^{\ast}/
\kappa$ is
different from
$a^{\ast}$ because $\kappa_{well} \equiv \kappa_{GaAs} \ne  \kappa_{barrier}
\equiv
\kappa_{Al_{x}Ga_{1-x}As}$ in the double quantum well system. 
(The quantitative correction arising from this difference is very
small since $a^{*}=98.3 \mbox{\AA}$ and $a=95.6 \mbox{\AA}$.)

\indent Before presenting our results we make some brief remarks about
Eqs.(\ref{eqn1})--(\ref{eqn5}).
First, we note that there is some arbitrariness in our definitions of the
symmetry-broken phases $\mbox{S}_{1}$, $\mbox{A}_{0}$, and $\mbox{A}_{1}$.
In particular, each spin polarized phase ($\mbox{S}_{1},\mbox{A}_{1}$)
is characterized by $|m|=1$, and therefore we could choose, for example,
for $\mbox{S}_{1}$: $m_{1}=m_{2}=1$ and for $\mbox{A}_{1}$: $m_{1}=0$,
$m_{2}=-1$. It is obvious that this arbitrariness does not affect energetics
and the calculated phase diagram, and is just the usual
arbitrariness of the order
parameter
in the broken symmetry phase. Second, we point out that the symmetry broken
phases $\mbox{S}_{1}$, $\mbox{A}_{0}$, and $\mbox{A}_{1}$ are completely
spin/layer polarized phases where the symmetry-broken order parameters
($m_{i}$,$n_{i}$) take on their maximum (in magnitude) values allowed
({\em i.e.} $|m_{i}|=1$, $|n_{i}|=2n$). In general, \underline{partial}
spin/layer polarization phases where, for example, $|m_{1}| \ne |m_{2}|$ with
$|m_{1}|$, $|m_{2}| \ne 0$ or $|n_{1}| \ne |n_{2}|$ with $|n_{1}|$,
$|n_{2}| \ne 0$ are allowed, but our energetic calculations have not found any
of these partial polarization phases to be global energy minima for any values
of $n$-$d$ parameters. We, therefore, believe that within our model partial
spin/layer polarization phases are not stable ground states for any
values of the parameters. 
   We obtain our restricted HFA phase diagram by minimizing
$E^{\rm HF}(m_{1},m_{2};n_{1},n_{2})$ with respect to the order parameters
$(m_{1},m_{2};n_{1},n_{2})$ for each value of the system parameters
$(d,n)$. Each $(d,n)$ point provides a unique set of $(m_{1},m_{2};n_{1},
n_{2})$ which minimizes the HFA energy, and thus a complete
phase diagram in $(d,n)$ parameter space is obtained, as shown in Fig.
\ref{II}.

   In Fig. \ref{II} we show the results of our simple HFA phase
diagram which allows for only three phases $\mbox{S}_{0}$, $\mbox{S}_{1}$,
and $\mbox{A}_{1}$ with the ${A}_{0}$ phase \underline{not}
stable at any values of the system parameters. Our calculated phase boundary
(triangles in Fig. 2) between the high density ({\em i.e.} low $r_{s}$)
paramagnetic
bilayer phase ($\mbox{S}_{0}$) and the low density ferromagnetic bilayer
phase ($\mbox{S}_{1}$) occurs at a fixed $r_{s}=2.011$ (by contrast, the
corresponding three dimensional HFA ferromagnetic instability occurs
at $r_{s}=5.45$) for all values of $d$ because this is just the two
dimensional HFA instability to the formation of a ferromagnetic
phase in which the Hartree energy does not play any role (note that interlayer
correlations are being neglected in our approximation). At still lower
(higher) density ($r_{s}$) there is a transition (the phase boundary
marked by squares) from
the bilayer ferromagnetic phase ($\mbox{S}_{1}$) to the monolayer
ferromagnetic phase ($\mbox{A}_{1}$) in Fig. \ref{II} --- this transition moves
to lower densities (higher $r_{s}$) as the interlayer separation $d$
increases because of the higher cost in Hartree energy. Also shown in
Fig. \ref{II} are three lines. The lowest line (the dotted line) is the $r_{s}
=d/a^{\ast}$ line, which distinguishes approximately the regime of the
average intralayer inter-electron separation ({\em i.e.} $r_{s}$) being
larger/smaller (the regime above/below the dashed line) than the
average interlayer
inter-electron separation ({\em i.e.} $d/a^{\ast}$) in dimensionless units.
The line with 
three dots and a dash in Fig. \ref{II} is the phase
boundary
between the paramagnetic bilayer phase (below this line) and the paramagnetic
monolayer phase (above this line), which is obtained if the spin 
polarization is ignored.
We also show in Fig.
\ref{II} by a solid
line the calculated phase boundary between the bilayer (below the line) and
the monolayer (above the line) phases for \underline{spinless} electron
systems, where, by definition, the ferromagnetic spin polarized phases do not
exist. Not
surprisingly, the phase boundary (solid line) for spinless fermions coincides
with the phase boundary (squares) separating the bilayer ($\mbox{S}_{1}$)
and the monolayer ($\mbox{A}_{1}$) spin polarized ferromagnetic phases
because the spin degree of freedom is frozen in the ($\mbox{S}_{1}$,
$\mbox{A}_{1}$) spin polarized phases.

       From Fig. \ref{II} we conclude that within the restricted HFA
any charge transfer instability between bilayer and monolayer phases (we
emphasize that fractional layer/spin occupancy states are not found to be
ground states for any values of $n$ and $d$) in double quantum well systems
must
necessarily be preceded by a ferromagnetic phase transition and the charge
transfer instability (the squares in Fig. \ref{II}) is the $\mbox{S}_{1}
\rightarrow
\mbox{A}_{1}$ transition. 
Inclusion of interlayer correlation and tunneling effects
should favor the bilayer phase over the monolayer phase, but our 
results show that, within the restricted HFA,
there is indeed a low density bilayer to monolayer
charge transfer instability.

\section{Unrestricted Hartree-Fock Theory}
\label{uhft}
In this section, we study the double-layer spin-${1\over2}$
interacting electron gas in the unrestricted HFA.
We show that the predicted bilayer to monolayer charge
transfer transition at low electron densities
(see the previous
section) is an artifact of the restricted HFA 
resulting from treating the layer-index as
a classical Ising-like variable.
Treating the layer degree of freedom in a fully quantum mechanical
fashion,
we show in this section that,
within the unrestricted HFA,
the charge-transferred monolayer
states are energetically unfavorable under any condition.
This generic conclusion regarding the nonexistence of a charge transfer
instability is
rigorously true in the HFA. 

For convenience, we adapt the pseudospin description
\cite{MACD1,MOON1,HALPERIN}
for the layer degree of freedom, where $\sigma_z=\pm1$ represent
the electronic states localized in the left and right layers, respectively,
and $\sigma_x=\pm1$ represent the symmetric and antisymmetric
states formed by the linear combinations of the left/right or $\sigma_{z}=\pm1$
eigenstates, respectively.  In this language, population of all the electrons
in a single layer corresponds to the pseudospin polarization
in the $\hat{z}$ direction, and population of all the electrons
in the symmetric state corresponds to the pseudospin polarization
in the $\hat{x}$ direction. Thus, these two states (monolayer occupancy and
symmetric state occupancy) are just pseudospin rotations of each other.
In the HFA, the ground state of low density electron
systems tend to have complete spin and pseudospin polarizations
in order to optimize the exchange energy.
Since the Hamiltonian of an electron gas is spin-rotationally invariant
(SU(2) symmetry),
the energy of the system does not depend on
the orientation of the spin polarization.
The Hamiltonian of the double-layer system
is, however, pseudospin-dependent (U(1) symmetry), so the energy
depends on the orientation of the pseudospin polarization.
As we will see shortly, the
suggested bilayer to monolayer charge transfer
instability is an artifact arising from the classical restriction
of the pseudospin polarization to the $\hat{z}$ direction. The pseudospin
rotated $\sigma_x-$polarized state necessarily has a lower energy than the
monolayer
occupancy $\sigma_z-$polarized state because there is no Hartree energy cost
associated with charge transfer in the $\sigma_x-$polarized state.

To study the dependence of the ground state properties on
the orientation of the pseudospin polarization,
we define the following orthonormal base
in the pseudospin space
\begin{eqnarray}
\label{equ:l1}
|\xi\rangle=&&\alpha|L\rangle+\beta|R\rangle,\nonumber \\
|\overline{\xi}\rangle=&&\beta|L\rangle-\alpha|R\rangle,
\end{eqnarray}
where $|L\rangle$ and $|R\rangle$ represent the $\sigma_z=\pm1$ electronic
states
localized in the left and right layers, respective,
$\alpha$ and $\beta$,
with $|\alpha|^2+|\beta|^2=1$,
are the pseudospin rotation parameters determining
a direction in the pseudospin space.
Because of the symmetry of the system,
we need only to consider the case where
both $\alpha$ and $\beta$ are real numbers
with $1\geq\alpha\geq\beta\geq0$. Our unrestricted HFA
consists of doing the energy minimization with $\alpha$, $\beta$ as free
parameters (with the constraint $|\alpha|^2+|\beta|^2=1$)
whereas the earlier restricted HFA (section \ref{RHFT}) made 
the specific choice of $\alpha/\beta=1/0$ (or,$0/1$).

We will examine the dependence of the
HFA energy of the electron gas
on the orientation of the pseudospin polarization,
and compare it to that of an unpolarized ({\em i.e.} bilayer) state.
For definiteness, we assume that there is no interlayer tunneling,
since the effect of the interlayer tunneling is always to oppose the charge
transfer instability.
The ground state of the spin and pseudospin unpolarized
phase (the $S_0$ phase of sec. \ref{RHFT}) is given by
$|S_0\rangle=\Pi_k C^\dagger_{k\xi\uparrow}C^\dagger_{k\xi\downarrow}
C^\dagger_{k\overline{\xi}\uparrow}C^\dagger_{k\overline{\xi}\downarrow}
|0\rangle$, where $C^\dagger_{k\xi s}$ ($C_{k\xi s}$)
is the creation (annihilation) operator for an electron
with momentum $k$, pseudospin $\xi$, and spin $s$,
and $|0\rangle$ is the vacuum state.
The corresponding HFA energy of the electron gas is
\begin{equation}
E^{\rm HF}_{S_0}=\left({1\over r_s^2}-{8\sqrt{2}\over3\pi r_s}\right)
{ne^2\over a^*},
\label{equ:l2}
\end{equation}
where $r_s$ is related to the electron density through
$n=1/(\pi a^{*2}r_s^2)$.
The ground state of the spin polarized but pseudospin unpolarized
phase (the $S_1$ phase) is given by
$|S_1\rangle=\Pi_k C^\dagger_{k\xi\uparrow}
C^\dagger_{k\overline{\xi}\uparrow}|0\rangle$.
The corresponding HFA energy of the electron gas is
\begin{equation}
E^{\rm HF}_{S_1}=\left({2\over r_s^2}-{16\over3\pi r_s}\right)
{ne^2\over a^*}.
\label{equ:l3}
\end{equation}
The ground state of the spin and pseudospin polarized phase is given
by $|P_\xi\rangle=\Pi_k C^\dagger_{k\xi\uparrow}|0\rangle$.
The corresponding HFA energy of the electron gas is
\begin{equation}
E^{\rm HF}_{P_\xi}(\alpha,\beta)=\left[{4\over
r_s^2}+{2d(\alpha^2-\beta^2)^2\over
ar_s^2}+\alpha^2\beta^2[I(0,r_s)-I(d,r_s)]\right]
{ne^2\over a^*},
\label{equ:l4}
\end{equation}
where
\begin{equation}
I(d,r_s)={4a^*\over d\pi}\int_0^1dxx\int_0^\pi d\theta
\left[1-e^{-{2\sqrt{2}d\over a^*r_s}
\left(\sqrt{1-x^2\sin^2\theta}-x\cos\theta
\right)}\right].
\label{equ:l5}
\end{equation}.

As shown in Eq. (\ref{equ:l4}), the energy of
the spin and pseudospin polarized state explicitly depends on the orientation
of the pseudospin polarization because the Coulomb interaction is layer index
dependent.
It is straightforward to show that the minimum of
$E^{\rm HF}_{{\rm P}_\xi}(\alpha,\beta)$ occurs
when $\alpha=\beta=1/\sqrt{2}$, {\it i.e.} when all the electrons
reside in the symmetric state,
and the maximum of
$E^{\rm HF}_{{\rm P}_\xi}(\alpha,\beta)$ occurs
when $\alpha=1$ and $\beta=0$,
{\it i.e.} when all the electrons reside in a single layer.
Thus, in the pseudospin space the monolayer occupancy phase is, in fact, an
energy maximum for the possible pseudospin polarized states of the system.
While both the symmetric state and the monolayer state
optimize the exchange interaction energy
by having complete pseudospin polarization,
the symmetric state has on the average equal electron densities
in the two layers and hence pays no cost in
the static charging energy (the Hartree energy).
The optimization of the exchange energy due to the pseudospin polarization
is somewhat larger in the monolayer state than in the symmetric state because
the intralayer Coulomb interaction is larger than the interlayer Coulomb
interaction,
but this difference is small compared with the Hartree energy cost
for any values of the layer separation and electron density.
Hence, the symmetric ($\sigma_x-$polarized) state is always energetically
favored
over the monolayer ($\sigma_z-$polarized) state.
Note, however, that if the electrons were not charged objects so that 
there was no Coulomb charging energy involved
in the charge transfer instability, then exchange energy by itself 
is better optimized by the $\sigma_z-$polarization and the 
monolayer state would be stable at low density.

In Fig.\ref{lf1}, we show the calculated HFA energies
of the double-layer spin-${1\over2}$ interacting electron gas
in the spin and pseudospin
unpolarized state ($E^{\rm HF}_{{\rm S}_0}$),
in the spin polarized but
pseudospin unpolarized
state ($E^{\rm HF}_{{\rm S}_1}$),
in the spin polarized symmetric state
$\left[E^{\rm HF}_{\rm SP-SY}=
E^{\rm HF}_{{\rm P}_\xi}\left(1/\sqrt{2},1/\sqrt{2}
\right)\right]$,
and in the spin polarized monolayer state
$\left[E^{\rm HF}_{\rm SP-MO}=E^{\rm HF}_{{\rm P}_\xi}(1,0)\right]$
as functions of the layer separation $d$ at different electron densities.
As mentioned above, $E^{\rm HF}_{\rm SP-MO}$ is always larger than
$E^{\rm HF}_{\rm SP-SY}$, hence, the bilayer to monolayer
charge transfer transition can never occur under any conditions.
In the HFA,
the energies of the pseudospin unpolarized states
are independent of the layer separation because there is no
interlayer direct or exchange interaction.
On the other hand, the energies of the pseudospin polarized states
are monotonically increasing functions of the layer separation.
The ground state of the spin-${1\over2}$ double-layer
system is found to be the spin and pseudospin unpolarized (paramagnetic
bilayer) state
at high electron densities, the spin polarized but pseudospin
unpolarized (ferromagnetic bilayer) state at intermediate densities,
and the spin-polarized symmetric (ferromagnetic $\sigma_x-$pseudospin-polarized
) state at low densities.
The calculated unrestricted HFA phase digram
is shown in Fig.\ref{lf2}.
There are three stable phases: the spin and pseudospin
unpolarized phase ($S_0$ phase),
the spin polarized but pseudospin unpolarized phase ($S_1$ phase ), and the
new spontaneously interlayer phase-coherent spin-polarized symmetric
phase (SP-SY phase).
This phase diagram is similar to that of Fig. \ref{II},
except for one
fundamental difference---the phase which
exists at low densities and small layer separations
in Fig. \ref{lf2} is not the charge transferred monolayer phase,
but the pseudospin-rotated spin polarized symmetric phase where electrons
on the average equally
populate both layers.
Inclusion of interlayer tunneling 
further reduces the energy of the SP-SY phase, making it even more
energetically favored over the monolayer $A_1$ phase. 
The spin polarization transition $S_0\rightarrow S_1$ is not affected 
by tunneling since the tunneling Hamiltonian
is spin independent.

Our focus here is on the interesting spontaneous interlayer phase-coherent
transition even in the absence of 
any tunneling energy.
We emphasize that the transition from the $\mbox{S}_{1}$ phase to the SP-SY
phase in Fig. \ref{lf2} is a true phase transition involving the
spontaneous breaking of the pseudospin symmetry because it happens
\underline{even in the absence of any interlayer tunneling}.
Without any interlayer tunneling ({\em i.e.} no spatial wavefunction
overlap between the layers) the layer index is conserved in the Hamiltonian,
and therefore the symmetric state, which is the even linear combination of the
$|L,R \rangle$ eigenstates, cannot be an eigenstate of the Hamiltonian unless
there is a spontaneous breaking of the layer symmetry. An equivalent
statement is that in the absence of tunneling one expects the ground state
to be an eigenstate of the $z$-component of pseudospin $\sigma_z$,
\underline{not} an eigenstate of $\sigma_x$ as the symmetric state is.
This is simply because the system Hamiltonian commutes (does not commute)
with the $\sigma_z$ ($\sigma_x$) operator.
In the presence of wavefunction overlap ({\em i.e.} when interlayer tunneling
is allowed) between the layers, the symmetric state is an allowed eigenstate
of the system and is trivially the ground state of the noninteracting double
quantum well structure. What is extremely interesting is our finding that even
in the strict absence of interlayer tunneling, exchange interaction can drive
the ground state of the system into the symmetric state ({\em i.e.} an
eigenstate of
$\sigma_x$) at low electron densities. This is a surprising result because in
the absence of tunneling all the terms in the Hamiltonian (the two dimensional
intralayer kinetic energy and the Coulomb interaction) conserve the layer index
of an electron whereas the symmetry broken ground state turns out to be a
coherent superposition of the electron being in the left and the right well
state.
In the strict absence of any tunneling, the low density SP-SY phase of
Fig. \ref{lf2} is an example of spontaneous interlayer phase coherence.
\cite{MOON1}
We note that the layer density is not a good quantum number in the SP-SY phase
even though there is no interlayer tunneling in the system! 
Thus the U(1) symmetry of the double quantum well Hamiltonian 
(without any tunneling ) is broken spontaneously in the SP-SY phase. While
being extremely interesting theoretically, the practical aspects of this
spontaneous rotation in pseudospin space (from an eigenstate of $\sigma_z$
to a symmetry-broken eigenstate of $\sigma_x$) remain unclear because
in the presence of any finite tunneling, 
the system should indeed be an eigenstate of $\sigma_x$. It is
certainly possible to make
double quantum well samples of high $r_{s}$ and low $d$, which also have
negligible interlayer tunneling (by having a very high potential barrier
between the
two layers). Our prediction is that such a system, if it is indeed in the SP-SY
phase of Fig. \ref{lf2}, would behave as if it is in the (tunneling induced)
symmetric state even though the actual tunneling matrix element is zero.
The situation is analogous to a quantum Hall system \cite{MOON1} at the 
filling factor of one, 
where it is believed that even in the absence of any Zeeman splitting
there will be a spontaneous exchange-driven spin polarization transition.
In our case, we have a spontaneous exchange-driven pseudospin polarization.

\section{Local Density Approximation}
\label{LDA}

\indent There have been several recent experimental studies
\cite{KAT1,YING,PATEL}
searching for the charge transfer instability.
All these studies involve applying an external electric field
(along the $z$ direction) to continuously tune electron densities and then to
measure layer electron densities via SdH oscillations.
The experimental work
involves \cite{KAT1,YING,PATEL} a GaAs-$\mbox{Al}_{x} \mbox{Ga}_{1-x}$As double
quantum
well structure (with an AlAs barrier layer) with an applied bias voltage
between a front gate
and the quantum wells. The action of this gate is to produce an electric field
that is external to the device and to draw electrons from the quantum wells
thereby lowering the total electron density of the system. We study this system
in the presence of an external electric
field (and also in zero external field) using both the self-consistent local
density
approximation (LDA) and the self-consistent local spin density approximation
(LSDA) to
determine the
electron density in each well as well as the polarization 
state of the electron
gas in each well, which can then be compared with the experimental results.

  The basic idea behind the LDA for the spin unpolarized case is to self
consistently solve the coupled Poisson equation and the one-dimensional
Schr\"{o}dinger-like Kohn-Sham equation \cite{KOHN} in order
to obtain
the ground state electron density of the
quantum well.
The LSDA is used to explore the possibility of a
spontaneous spin polarization transition in the system.
The LSDA is similar in spirit
\cite{AC,BH} to the aforementioned LDA. The major
differences are that there are two Kohn-Sham equations (one for each spin
component) that need to be solved in LSDA and that the exchange-correlation
potential
now depends \cite{BH} on both the electron density and the spin polarization
of the electron gas. 
For the exchange-correlation potentials, we use the
parameterization of Ceperley and Alder \cite{AC} 
for spin unpolarized or completely polarized and an
interpolation formula
due to von Barth and Hedin \cite{BH} for partial polarizations.
Details of LDA \cite{EISEN} and LSDA \cite{RTD} calculations 
for double-layer systems can be found in the literature.

Because the spin polarization of the final state can be affected by
numerical inaccuracies, we perform the calculation using two very different
initial values of polarization.
One choice is a starting polarization that is small ($10 \%$) and the
other is a starting polarization that is large (90\%). If both
choices lead to a polarized final state then we say that state is polarized.
If only one choice (say the 90\% initial polarization) leads to a
polarized final state, then we assume that the final result is affected by
numerical inaccuracies and is uncertain.
We follow the procedure of Eisenstein {\em et. al.} \cite{EISEN}
in allowing for the interlayer charge transfer in the presence of the
external bias voltage within the LDA and LSDA.
The external electric field is generated by adding additional charge
to the top-most donor impurity sheet while maintaining overall charge
neutrality between the donor sheets and the quantum wells.

\indent We reproduce the results of Ying {\em et. al.} \cite{YING} for
two different double quantum
well structures in the presence of an external electric field in Fig.
\ref{VII}(a), (b), and (c) finding good
quantitative agreement between their results and ours. In Fig. 
\ref{VII}(a) and (b)
we calculate the front and back layer densities within the Hartree
approximation (dashed line) and the LDA (solid line). The linear behavior of
the electron densities within the Hartree approximation is expected as the
electrons in the
front layer attempt to screen the electrons in the back layer from the electric
field. However, within the LDA, the density of the front layer decreases more
quickly than in the Hartree approximation and the back layer density
actually increases with increasing electric field. In Fig. 
\ref{VII}(c), we show the
results of the LSDA (three dots and a dash) showing that even with the
possibility
of spin polarized states, the density of the back layer increases. This
increase in the back layer density is a manifestation of the
exchange-correlation
induced
``negative compressibility'' effect \cite{EISEN} which leads to a nonlinear
layer depopulation in the external voltage.

\indent We also perform the LSDA calculation on the same double quantum well
structure as in Fig. \ref{VII}(c) without an external electric field
to calculate the ground state phase diagram of the system. Our phase
diagram, in electron density--layer separation space shown in
Fig. \ref{VIII}, indicates
that as the total density of the system is decreased, charge is not
transferred from one well to the other, but instead, we find spin polarized
ferromagnetic states [our $\mbox{S}_{1}$ states of Fig. \ref{II}] for the
electron gas in both wells. This establishes that although
there is a net interlayer charge transfer 
\underline{in the presence of an external
electric field}, it is not the exchange-driven spontaneous
bilayer-to-monolayer
charge transfer
instability.
We note that our LSDA phase diagram shown in Fig.\ref{VIII}
is qualitatively similar to the HFA phase diagram (e.g., 
Fig.\ref{II}) with two important differences: (1) The monolayer
phases are not present, and (2) the ferromagnetic transition occurs at a
somewhat higher (lower) $r_s$ (density) value, which is expected because the realistic LSDA calculation includes effects of finite well widths etc.
and includes correlation effects.

\indent Our LDA and LSDA calculations establish that the external electric
field induced interlayer charge transfer experiments can be quantitatively
understood as ``negative compressibility'' \cite{EISEN}
effects. This point, in
fact, has already been made in some of the experimental publications
\cite{YING,PATEL} where good agreement between the experimental data and the
LDA
calculations was shown to exist. Our predicted
LSDA calculation based 
ferromagnetic spin polarization transition (Fig. \ref{VIII})
should occur at substantially lower densities than the experimental
densities utilized in the existing literature.\cite{KAT1,YING,PATEL}
Our predicted densities for the spin polarization transition 
in semiconductor double well systems should, however, be accessible,
particularly in hole-doped samples,\cite{SIVAN}
where large effective $r_s$ ($\ge25$) values have been recently 
achieved experimentally.

\indent We emphasize that the experimental measurements carried out in the
presence of an external electric field have little to do
with the theoretical issue of a bilayer to monolayer charge transfer
instability
because the application of the external electric field necessarily destroys
the layer symmetry in the problem and the issue of a spontaneous symmetry
breaking (phase) transition or exchange instability becomes irrelevant. The
situation is analogous (but \underline{not} identical) to a magnetic
transition in the presence of an external magnetic field, which is not a phase
transition in any sense because there is an applied symmetry breaking field. As
mentioned
before, the depopulation of subbands in the presence of an applied gate voltage
is nonlinear due to exchange-correlation effects, and the so-called
\cite{EISEN} exchange-correlation induced ``negative compressibility'' effect
is the
cause of the ``unexpected'' bump seen in the experimental results (see our
Fig. \ref{VII} and refs. \onlinecite{KAT1,YING,PATEL}), which is quantitatively
explained by the LDA/LSDA calculations. 

\section{Conclusion}

\indent We have obtained four new theoretical results in this paper:

\begin{enumerate}
\item We have shown that there cannot be any exchange driven bilayer to
monolayer charge transfer instability in semiconductor double quantum well
systems.

\item We have shown, within a mean field HFA and also
within a self-consistent LSDA theory, that there could be a ferromagnetic
spin polarization transition in a double quantum well system at low (but
accessible) densities.

\item We have shown that within a mean field unrestricted HFA
there is a quantum phase transition in a double quantum well system from a
(spin polarized) bilayer state to a (spin polarized) interlayer phase-coherent
symmetric state at low electron densities even in the absence of
any interlayer electron tunneling---in the symmetric state the electron
density in each layer develops spontaneous quantum fluctuations even though
there is no overlap between the layer 
wavefunctions in the absence of tunneling.

\item We have shown that the experimental measurements \cite{KAT1,YING,PATEL}
of
layer/subband charge densities in double quantum well systems as a function
of an applied external electric field can be understood quantitatively on
the basis of LDA/LSDA calculations 
as arising from the two dimensional ``negative compressibility '' effect.

\end{enumerate}

\indent Of these four results, obviously the most interesting are the
results (2) and (3) above, both of which are based on reasonable but
approximate theories. Our HFA phase diagrams
invariably show low density
ferromagnetic phases
where the electrons in each layer undergo a complete spin polarization
transition. Our numerical LSDA calculation (Fig. \ref{VIII}) also finds the
same result. While it is certainly possible (may even be likely) that our
mean field theory overestimates the density at which the ferromagnetic
transition occurs, we believe that at high (low) enough $r_{s}$ 
(density) the
semiconductor
double quantum well system does undergo a spin polarization transition.
It would be difficult experimentally to directly observe this ferromagnetic
transition because the actual spontaneous electronic 
magnetic moment associated
with the spin polarized 
two dimensional electrons is rather small, and would be 
difficult to measure because of the large (orders of magnitude larger)
background effect arising from the lattice. For the same reason, standard
thermodynamic measurements ({\em e.g.} heat capacity) of the ferromagnetic
phase transition may also be impossible. One possibility is to measure the two
dimensional Fermi momentum $k_{F}$ in each layer which will exhibit a jump
by a factor of $\sqrt{2}$ at the ferromagnetic transition. 
Transport measurements (in individual layers or 
in interlayer drag experiments) 
may
be useful in this respect because, in principle, such measurements
\cite{SIVAN,GRAM} are
capable of indirectly measuring $k_{F}$.
We believe transport \cite{SIVAN,GRAM}
and capacitance \cite{EISEN} spectroscopies should show observable 
structures at the spin polarization transition in density 
sweep experiments.

\indent Our most interesting theoretical finding is the possibility of a
low density quantum phase transition from a bilayer state to a coherent
interlayer symmetric state [{\em cf.} Fig. \ref{lf2}], which happens even in
the
absence of any interlayer tunneling. Within the HFA we believe
the existence of this phase transition to be rigorous. We speculate that this
phase transition would exist even when correlation effects 
are included in the theory because correlation should affect the
bilayer and the symmetric phase more or less equivalently. 
While being very interesting theoretically in
its own right, a definitive experimental observation of this exchange driven
bilayer to symmetric phase transition [{\em cf.} Fig. \ref{lf2}] in
semiconductor double quantum well systems would be difficult for a number of
reasons.
First, very low electron 
density ($\sim 10^{9}$$\mbox{cm}^{-2}$) and
low disorder double quantum well samples will be needed with rather large AlAs
potential barriers to suppress tunneling. This is currently beyond the reach of
MBE growth techniques for electron doped samples. It is, however, possible to make p-doped hole
samples with very large 
effective $r_{s}$ ($\stackrel{\sim}{>}25$) values \cite{SIVAN},
which may be more
suitable for observing our predicted 
spontaneous phase coherent transition. Even if the desired samples
are produced, few experiments 
(short of actual thermodynamic measurements which
can look at specific heat anomalies at the phase transition) can actually
distinguish between bilayer and symmetric states, because both states have
the same \underline{average} layer electron densities --- in one case 
(bilayer)
the layer electron density is an exact quantum number with no fluctuations
while in the other case (symmetric) the layer electron density has quantum
fluctuations and is not conserved. It seems that the interesting quantum phase
transition shown in Fig. \ref{lf2} may remain only a tantalizing theoretical
possibility in the near future.

A natural question arises about the nature of the quantum phase transition
involved in the spontaneously breaking of the U(1) pseudospin symmetry
in going from the bilayer $S_1$ phase to the interlayer 
phase coherent SP-SY phase in Fig.\ref{lf2}. 
Note that a similar phase transition has earlier been discussed in the 
literature \cite{MOON1} in the context of the quantum Hall effect phenomena 
in bilayer systems where it has been argued that the U(1) symmetry of a double
quantum well system in the absence of tunneling is spontaneously broken
at Landau level filling factor of one (and possibly at other 
filling factors as well).
We find that there is nothing special about the quantum Hall situation
in this context, and in fact as we show in this paper, 
a zero field exchange-induced spontaneous breaking of the U(1) 
symmetry is indeed possible at low densities. The theoretical 
phase diagram (Fig.\ref{lf2}) in the zero field situation is, 
in fact, richer because there are two tuning parameters ($r_s$ and $d$)
controlling the phase transition (whereas in the quantum Hall case $d$ 
is the only tuning parameter).
Experimentally, of course, the situation is much more easily realized
in the quantum Hall situation because it is much easier to 
obtain a Landau level filling factor of one than an $r_s$ of ten. 
We speculate \cite{rk3}
that the nature of the U(1) pseudospin symmetry breaking
phase transition in the zero magnetic field case is similar 
to that in the finite field quantum Hall situation,\cite{MOON1}
even though further investigation of this issue in the zero 
field case is clearly warranted. 
While all the implications \cite{rk3} of such a transition
in our zero field case still remain to be worked out, it
is likely that there is (at least the possibility of ) an 
interesting finite temperature transition.  In this context it may be 
worthwhile to point out that spontaneous interlayer phase coherence has been 
argued \cite{vig} to lead to interesting and observable effects
in interlayer drag experiments.\cite{GRAM}
We believe that such effects \cite{vig} would show up in the
SP-SY phase as well and may be a way of identifying the new phase.
More work is clearly needed in establishing the properties 
of the SP-SY phase and in elucidating the nature of the U(1) 
pseudospin-symmetry-breaking phase transition.

\indent One of the authors (Sankar Das Sarma) acknowledges a helpful
conversation
with Professor B. I. Halperin. This work is supported by the United States
Office of Naval Research (U.S.-O.N.R.).

\begin{figure}[p]
\caption{The definitions of the four phases, $\mbox{S}_{0}$, $\mbox{S}_{1}$,
$\mbox{A}_{0}$, and $\mbox{A}_{1}$ used in the the Mean Field Phase
Diagrams.}
\label{I}
\end{figure}

\begin{figure}[p]
\caption{Phase diagram for the restricted Hartree-Fock theory.
The dotted line is the $r_{s}=d/a^{*}$ line.
The line with the three dots and a dash is the phase boundary of ref.
\protect\onlinecite{RW}.
The solid line is the phase boundary for spinless fermions.
$r_{s}= 1/a^{*} (\pi n)^{1/2}= 5.73759 \times 10^{5}/(n)^{1/2}$ for $n$
in units of $1/cm^{2}$.}
\label{II}
\end{figure}

\begin{figure}
\caption{
The Hartree-Fock energies
of a double-layer spin-${1\over2}$ interacting electron gas
in the totally unpolarized state ($E^{\rm HF}_{{\rm S}_0}$),
in the spin polarized but pseudospin unpolarized state
($E^{\rm HF}_{{\rm S}_1}$), in the spin polarized symmetric
state ($E^{\rm HF}_{\rm SP-SY}$), and in the spin polarized monolayer
state ($E^{\rm HF}_{\rm SP-MO}$) at different electron densities:
(a)$r_s=(2)^{1/2}$; (b)$r_s=2(2)^{1/2}$; (c) $r_s=4(2)^{1/2}$.
}

\label{lf1}
\end{figure}

\begin{figure}
\caption{
Phase diagram of a double-layer spin-${1\over2}$ interacting electron gas
in the Hartree-Fock approximation.  There are three stable phases:
the totally unpolarized phase ($S_0$ phase),
the spin polarized but pseudospin unpolarized phase ($S_1$ phase),
and the spin polarized symmetric phase (SP-SY phase). The charge transferred
monolayer phase ($\mbox{A}_{0}$ or $\mbox{A}_{1}$ of Fig. 1) is not found to be
a
stable phase for any values of ($r_{s},d$).
}
\label{lf2}
\end{figure}

\begin{figure}
\caption{
(a) Plot of layer electron density versus total electron density for a double
quantum well structure from ref. \protect\onlinecite{YING} with a barrier width
of $14 \mbox{\AA}$ and well widths of $180 \mbox{\AA}$.
The dashed line is the Hartree approximation. The solid line is the LDA.
(b) Plot of layer electron density versus total electron density for a double
quantum well structure from ref. \protect\onlinecite{YING} with a barrier width
of $70 \mbox{\AA}$ and well widths of $150 \mbox{\AA}$.
The dashed line is the Hartree approximation. The solid line is the LDA.
(c) Plot of layer electron density versus total electron density for the double
quantum well structure in (b). The dashed line is the Hartree approximation.
The solid line is the LDA calculation. The line with three dots and a dash is
the LSDA
calculation.
}
\label{VII}
\end{figure}

\begin{figure}
\caption{
Phase diagram of the double quantum well structure in Fig. \protect\ref{VII}
(b) within the
LSDA without an external electric field. The lower (upper) line corresponds
to an initial spin polarization of $10 \%$ ($90 \%$) in the LSDA calculation.
The region between these two lines is comprised of points for which the
final spin polarization is dependent on the initial spin polarization used
in the LSDA calculation.
}
\label{VIII}
\end{figure}


\begin{references}
\bibitem{rk3}
The spontaneous breaking of the $U(1)$ symmetry in the SP-SY phase corresponds
to the classical $X$-$Y$ model.
It is, therefore, natural to speculate that there is a finite temperature
Kosterlitz-Thouless transition in the SP-SY phase at a
critical temperature $T_{KT}$, 
with interlayer phase coherence of the SP-SY phase destroyed above $T_{KT}$.
Below $T_{KT}$, 
the SP-SY phase 
should support a Goldstone mode with vanishing energy at long wavelengths. 
We believe that many details of our zero field phase transition should 
correspond to the finite field quantum Hall situation discussed 
in ref. \protect\onlinecite{MOON1}.
\end{references}
\end{document}